\documentstyle[prl,draft,aps]{revtex}

\input epsf

\begin{document}
\twocolumn[\hsize\textwidth\columnwidth\hsize\csname@twocolumnfalse\endcsname

\def\lesssim{\mathrel{\hbox{\lower1ex\hbox{\rlap{$\sim$}\raise1ex\hbox{$<$}}}}}
\title{Magnetic character of the empty density of states 
in uranium compounds from X-ray magnetic circurlar dichroism}
\author{P. Dalmas de R\'eotier and A.~Yaouanc}
\address{Commissariat \`a l'Energie Atomique\\ D\'epartement de Recherche 
Fondamentale sur la Mati\`ere Condens\'ee\\ F-38054 Grenoble Cedex 9, 
France}

\date{\today} 
\maketitle

\begin{abstract}

We present a discussion of published x-ray magnetic circular dichroism 
(XMCD) measurements performed at the uranium M$_{4,5}$ edges of metallic 
uranium compounds, focusing on the shape of the dichroic signal at 
the M$_{5}$ edge. 
A well resolved double lobe structure, comprised of a positive and negative 
peak, is sometimes observed. Out of the twelve metallic uranium compounds so 
far investigated by XMCD, six exhibit an intense double-lobe structure at 
the M$_{5}$ edge. This line shape gives information on the empty 
5$f$ magnetic density of states with angular quantum number $j=7/2$. 
Conclusions about the difference between these two families of compounds 
are given regarding the splitting of the $j=7/2$ band and the occupation 
among the different $m_{7/2}$ sublevels.

\end{abstract}

\pacs{PACS numbers: 75.30.Mb, 74.70.Tx, 78.70.Dm, 78.20.Ls}

]

\section{Introduction}

In recent years uranium compounds have been the subject of increasing
interest because their ground state exhibits a variety of physical
properties. They can be Pauli paramagnets or display an ordered magnetic
state. The electronic correlations can be very large as revealed by a strong 
heavy fermion character or Kondo effect \cite{Grewe91}. Even more surprising, 
at low temperature four uranium compounds are both superconductors and 
magnetic at ambient pressure \cite{Heffner96}. 

Although the 5$f$ electrons of metallic uranium compounds are more easily 
treated in a localized magnetism framework, their hybridization with the 
conduction and ligand electrons can not be neglected \cite{Brooks84}. 
Their ground state properties reflect the competition between at least four 
types of interactions of about the same strength: Coulomb
and exchange, crystal field, hybridization and spin-orbit coupling.  
Due to the complexity of the physics involved, a complete understanding 
of an uranium compound has not yet been achieved.

In comparison to the vast theoretical literature on the 5$f$ electronic 
properties, experimental microscopic information is scarce. It is
usually obtained from  photoemission, de Haas-Van Alphen measurements, neutron
scattering and muon spin spectroscopy. With the advent of third
generation synchrotron radiation sources, new experimental techniques have 
become available, such as x-ray magnetic circular dichroism (XMCD) 
\cite{Lovesey96}. This paper presents a discussion of published XMCD spectra 
recorded at the M$_{4,5}$ edges of uranium atoms in metallics uranium 
compounds. Our interest here is to compare the XMCD spectra with the purpose 
to find relations between their characteristics and the electronic structure 
of the compounds. 

\section{The x-ray magnetic circular dichroism data}

We recall that the XMCD technique consists of recording two spectra 
at the absorption edges of a spin-orbit split core state chosen to probe the 
electronic state of interest. The two spectra differ by the handedness of the 
circularly polarized light used to record them. Information about the $5f$ 
uranium states can be obtained by performing measurements at the uranium 
M$_{4,5}$ edges. 

\begin{figure}
\centerline{
\epsfxsize 15 cm
\epsfbox{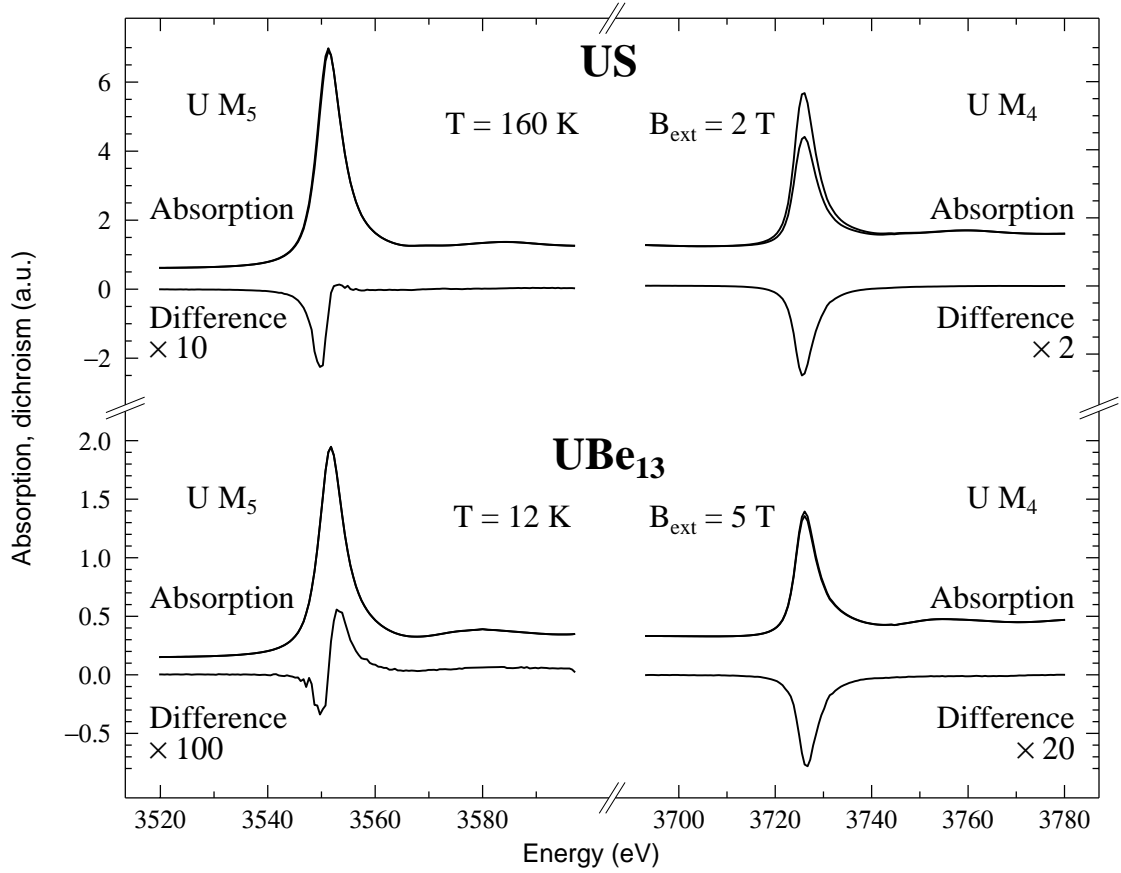}}
\caption{Examples of M$_{4,5}$ absorption spectra and corresponding 
dichroism \protect \cite{Dalmas99}.
The dichroic spectra are obtained by simple difference 
of the absorption spectra without any further data manipulation.}
\label{Examples}
\end{figure}

The XMCD technique is known to provide information on the orbital and 
spin magnetic moments through the use of sum rules which only involve
the integrated intensity of the absorption and dichroic spectra, {\sl i.e.}
they are independent of the shape of the XMCD spectra. However,  
measurements of twelve uranium compounds show that in fact the shape 
at the M$_{5}$ edge depends strongly on the compound. This is illustrated by 
two examples in Fig.~\ref{Examples}. The dichroism at the M$_{4}$ edge 
consists always of a single negative lobe that has no distinct structure. 
On the other hand, two strong lobes, a positive and a negative one, or a 
single lobe, can be  observed at the M$_{5}$ edge. In fact, a double-lobe 
structure is detected for UPd$_2$Al$_3$ \cite{Yaouanc98}, UBe$_{13}$ and 
UPt$_3$ \cite{Dalmas99}, U$_{0.3}$La$_{0.7}$S and U$_{0.4}$La$_{0.6}$S 
\cite{Bombardi01} and UGe$_2$ \cite{Kernavanois}. It does not
exist (or the second lobe is very small compared to the first one)
for the following six compounds: 
US \cite{Collins95,Dalmas99,Kernavanois01}, USb$_{0.5}$Te$_{0.5}$ 
\cite{Dalmas97}, UFe$_2$ \cite{Finazzi97}, URu$_2$Si$_2$ \cite{Yaouanc98},
URhAl \cite{Grange98} and U$_{0.6}$La$_{0.4}$S \cite{Bombardi01}. XMCD 
measurements on UNi$_2$Al$_3$ were performed but the signal intensity at 
the M$_5$ was too small to determine its shape \cite{Kernavanois00}.

\begin{figure}
\centerline{
\epsfbox{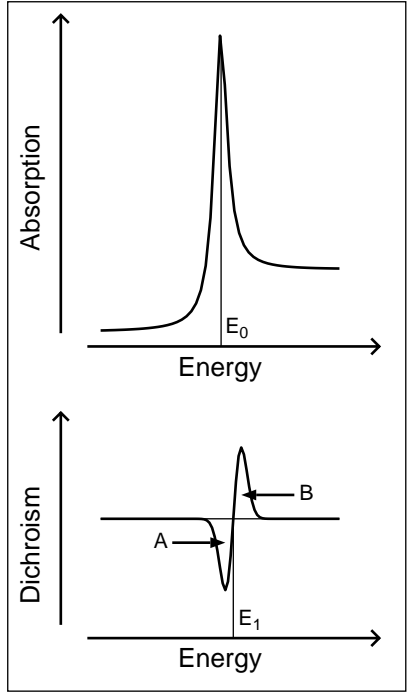}}
\caption{Drawing used to define the physical quantities needed to quantify 
the XMCD spectra at the M$_5$ edge. $E_0$ is the energy at 
which the absorption is maximum, $E_1$ the energy where the XMCD response 
vanishes between the two lobes, $A$ the algebraic area of the lobe located 
below $E_1$ and $B$ the algebraic area of the lobe appearing above $E_1$.} 
\label{definition}
\end{figure}

We quantify the shape of the XMCD spectra at the M$_5$ edge in 
Table~\ref{xray} by the ratio $R_a$ of the algebraic area of the two lobes 
and the relative energy $\Delta E$ at which the dichroism signal vanishes 
between them. The parameters introduced to characterize the XMCD response are 
defined in Fig.~\ref{definition}. When known, we also list in the table
the Sommerfeld coefficient which is a measure of the density
of states at the Fermi level and which is usually taken as a gauge of
electronic correlations between the conduction electrons, 
{\sl i.e.} larger is $\gamma$, stronger are these correlations. However, 
mechanisms other than electronic correlations can contribute to $\gamma$ 
such as an appreciable low lying crystal field energy level density. We note 
a remarkable correlation between the values of $\gamma$
and of both $R_a$ and $\Delta E$: when $\gamma$ is large, the two lobes
are clearly defined ($R_a  \ll 0$) and the point between the two lobes at 
which the XMCD spectrum vanishes is shifted to the left of the maximum of the 
absorption spectrum ($\Delta E \ll 0$). To understand these results we first 
discuss the physical meaning of the XMCD measurements.

\section{Implications on the electronic density of states}

Due to the selection rules for dipolar electronic transitions induced by light,
the structure at the M$_5$ edge of the dichroic resonance provides 
information on the magnetic character of the density of states (DOS) above the 
Fermi level \cite{Dalmas99,Shishidou99}.

We recall that the M$_4$ (M$_5$) edge corresponds to
$ 3d_{3/2} ( 3d_{5/2}) \rightarrow 5f$ transitions. The M$_4$ absorption 
signal is proportional to the number of $f_{5/2}$ holes, while the M$_5$ 
absorption signal depends primarily on the number of $f_{7/2}$ holes.
Since the XMCD technique uses circular polarized light, the
dichroism contains information about the magnetic character of the 
sublevels in the DOS.

We first use the atomic picture with the $jj$-coupling. For both 
M$_4$ and M$_5$ edges, within the conventions adopted, the unoccupied 
sublevels above the Fermi level with negative magnetic quantum number $m_j$ 
give a positive dichroism signal and those with positive $m_j$ values a 
negative one. Qualitatively, it is expected that the two or three $f$ 
electrons (the valency of the compounds under interest is expected to be 
between $+4$ and $+3$) mainly occupy the sublevels with negative $m_{5/2}$, 
i.e. $-5/2$, $-3/2$ and $-1/2$, so that most of the hole density is in 
sublevels with positive $m_{5/2}$ values.

With this background, we expect to observe an essentially negative dichroic 
signal at the M$_4$ edge. Such a feature has been systematically observed for 
all uranium compounds. 

We now consider the M$_5$ edge. The energy sequence of the $m_{7/2}$ sublevels
is opposite to the $m_{5/2}$ one: the negative $m_{7/2}$ 
sublevels are located at higher energy relative to the 
positive $m_{7/2}$ sublevels. This reflects the gain in energy due to the
alignment of the spin with the exchange field.
Perturbations such as hybridization mix levels with same $m_{j}$ values.
Crystal-field, Coulomb and exchange interactions mix levels of similar
$m_s$ and m$_l$ values. Since, within the $j=5/2$ levels, 
the negative $m_{5/2}$ sublevels are mostly occupied, it results that
the negative $m_{7/2}$ sublevels are also preferably electron occupied. 
Therefore, a relatively strong 
negative lobe at low energy is expected and an eventual weak positive lobe at 
high energy is possible. These predictions of the atomic model provide
a qualitative understanding of the results for the last six compounds of 
Table~\ref{xray}. In the case where the energy splitting of the $m_{7/2}$ 
sublevels is small compared to the intrinsic width of the electronic transition
(given by the core hole lifetime), the observation of a single lobe
remains, of course, expected.

Now, for the first six compounds of Table~\ref{xray}, we observe a 
redistribution of weight between the two lobes. This remarkable feature 
first implies that the
energy splitting within the $m_{7/2}$ sublevels is large. The high-energy 
lobe is even the more intense lobe for the first three compounds. This means 
that the negative $m_{7/2}$ sublevels are less electron occupied 
than expected from the atomic $jj$-coupling point of view, i.e. the 
density of empty negative $m_{7/2}$ sublevels is larger than expected.
 
We have already pointed out that the compounds with a large $R_a$ ratio, 
i.e. with two lobes, have their $E_1$ shifted to an energy smaller 
than $E_0$. We do not have yet an understanding of this effect. Interestingly,
we note that the majority of the spectral weight is at an energy smaller 
than $E_0$ for all the compounds, i.e. even for those exhibiting only one lobe.
Probably, the observation of the two lobes 
and of the shift of $E_1$ reflect the same physics. 

To understand the origin of the double-lobe, we may leave the $jj$-coupling 
scheme and work in the intermediate coupling scheme which allows a mixing of 
negative $m_{5/2}$ with positive $m_{7/2}$ sublevels. The breakdown of the 
$jj$-coupling scheme for the double-lobe compounds means that the 
hybridization, Coulomb and exchange and crystal field energies can no longer 
be taken as a perturbation relative to the 5$f$ spin-orbit interaction for 
these compounds. Indeed, it has been shown that, for example, a crystal field 
of 1 eV or larger can lead to a double-lobe structure \cite{Yaouanc98}. 
Although such a strong crystal field is not realistic, it shows that the 
breaking of the $jj$-coupling approximation leads to the double-lobe 
structure. 

Instead of starting from the atomic picture, one may use a band-like approach
as done by Shishidou and co-workers \cite{Shishidou99}. In fact, their 
computation for US yields a weak high-energy lobe as found experimentally.   
The band-like picture seems to be appropriate for the compounds with a 
double-lobe structure because of their large Sommerfeld coefficient which
means that their density at the Fermi level is large. However, no matter
the starting point of the description, {\sl i.e.} the atomic or the band
limit, our previous conclusion that the double-lobe structure is a signature
of a relative large density of empty negative $m_{7/2}$ sublevels is a 
robust result 
since it arises basically from the selection rules for x-ray induced atomic 
transitions.

\section{Conclusion}

Many experimental methods such as bulk techniques, neutron scattering
and $\mu$SR spectroscopy have suggested that strongly electronically 
correlated uranium compounds can be thought of as systems with two 
components: conduction electrons and local moments. UPd$_2$Al$_3$ 
\cite{Caspary93,Feyerherm94,Metoki98,Bernhoeft98,Sato01} seems to be the 
cleanest example with two localized $5f$ electrons per uranium in 
the U$^{4+}$ state responsible for the measured magnetic moment, while the
remaining $5f$ electron density is itinerant. UBe$_{13}$ \cite{Varma86} and 
UPt$_3$ \cite{Broholm90} are other two examples of the localized-itinerant
duality, although for these two compounds there is no magnetic moment 
(UBe$_{13}$) or it is very small (UPt$_3$). Recently, UGe$_2$ has been found 
to be similar to UPd$_2$Al$_3$ \cite{Yaouanc01}. The XMCD results suggest
that the low-energy lobe arises from the U$^{4+}$ state under the influence 
of the crystal field \cite{Yaouanc98} and the high-energy lobe is a signature
of the itinerant component. Within this picture, the latter component has a 
strong $m_{7/2}$ character, in particular for UPd$_2$Al$_3$. Clearly,
band structure computations are needed to test our suggestion.


\onecolumn
\widetext

\begin{table}
\caption{Specific heat and XMCD data for twelve uranium-based compounds. The 
values of the Sommerfeld parameter $\gamma$ are taken from the literature and 
the algebraic area ratio $R_a \equiv B/A$ and energy difference 
$\Delta E \equiv E_1 - E_0$ are estimated from published x-ray M$_5$ 
absorption and XMCD spectra. The parameters $A$, $B$, $E_0$ and $E_1$ are 
defined in Fig.~\ref{definition}. The compounds are classified by increasing
values of $\Delta E$ and an horizontal line is used to distinguish the 
compounds with negative or zero $\Delta E$ value from the other compounds.
We note that this order is also compatible, within the
error bars, to increasing values of $R_a$. A question mark in the $\gamma$ 
column means that for the given compound we are not aware of any published 
value for this parameter. 
An hyphen in the $\Delta E$ column means that the value is
irrelevant: when only one lobe is identified ($R_a$ = 0) neither $E_1$ nor
$\Delta E$ is defined.}
\begin{tabular}{lrrrrr}
Compound & \multicolumn{2}{c}{Specific heat data} & 
\multicolumn{3}{c}{XMCD data} \\
& $\gamma$(mJ.mol$^{-1}$.K$^{-2}$) 
& References & $R_a$ & $\Delta E$ (eV) & References \\ \hline \hline
UPd$_2$Al$_3$        & 145 & \protect \cite{Heffner96}   & $-24$ (10) 
& $-0.8$ (2) & \protect \cite{Yaouanc98}  \\
UBe$_{13}$           & 1100& \protect \cite{Heffner96}   & $-3.0$ (2) 
& $-0.5$ (2) & \protect \cite{Dalmas99}   \\
UPt$_3$              & 450 & \protect \cite{Heffner96}   & $-2.2$ (5) 
& $-0.2$ (2) & \protect \cite{Dalmas99}   \\
U$_{0.3}$La$_{0.7}$S & ?   & -         & $-0.5$ (2) 
& $-0.2$ (1) & \protect \cite{Bombardi01} \\
U$_{0.4}$La$_{0.6}$S & ?   & -         & $-0.4$ (2) 
& $-0.2$ (1) & \protect \cite{Bombardi01} \\
UGe$_2$              & 32  & \protect \cite{Huxley01}    & $-0.54$ (4)
&  0.1 (1) & \protect \cite{Kernavanois}\\
\hline 
U$_{0.6}$La$_{0.4}$S & 30  & \protect \cite{Bourdarot99} & $-0.06$ (1)
&  0.6 (2) & \protect  \cite{Bombardi01}\\
US                   & 23  & \protect \cite{Westrum68}   & $-0.04$ (1)
&  0.7 (2) & \protect \cite{Dalmas99,Kernavanois01} \\
UFe$_2$              & 45  & \protect \cite{Franse83}    & $-0.06$ (3)
&  0.8 (2) & \protect \cite{Finazzi97}  \\
USb$_{0.5}$Te$_{0.5}$& ?   & -                           & 0.0 (1)  
&  -       & \protect \cite{Dalmas97}   \\
URu$_2$Si$_2$        & 65  & \protect \cite{Heffner96}   & 0.00 (1) 
&  -       & \protect \cite{Yaouanc98}  \\
URhAl                & 76  & \protect \cite{Grange98}    & 0.00 (2) 
&  -       & \protect \cite{Grange98}   \\
\end{tabular}
\label{xray}
\end{table}

\end{document}